\documentclass[11pt,twoside]{article}

\usepackage{asp2006}
\usepackage{epsf}
\usepackage{graphicx}
\usepackage{lscape}

\def\kms{km~s$^{-1}$}
\def\etal{{\it et al.}}

\def\msun{$M_\odot$}

\begin{document}
\title{ALFALFA HI Content and Star Formation in Virgo Cluster Early-Type Dwarfs}   
\author{R. A. Koopmann$^1$, R. Giovanelli$^2$, M. P. Haynes$^2$, N. Brosch$^3$}   
\affil{$^1$Department of Physics and Astronomy, Union College,
  Schenectady, NY 12308\\ $^2$Center for Radiophysics and Space Research, Cornell University, Ithaca, NY 14853\\$^3$Wise Observatory and the Beverly and Raymond Sackler School of Physics and Astronomy, Tel Aviv University, Tel Aviv 69978, Israel}    

\begin{abstract} 

Early-type dwarfs dominate cluster environments, yet their
formation and evolutionary histories remain unclear. 
The ALFALFA (Arecibo Legacy Fast ALFA) blind survey is providing a 
census of HI in galaxies of all types in a range of environments.
Here we report on ALFALFA results for
Virgo Cluster early-type dwarfs between declinations of 4 and 16 degrees.  Less
than 2\% of the Virgo early-type dwarf population is detected, compared to
70-80\% of the Im/BCD dwarf population.
Most of the dwarfs detected in HI show
evidence for ongoing or recent star formation.  Early-type galaxies with
HI tend to be located in the outer regions of the cluster and to
be brighter.  Early-type dwarfs with HI may be
undergoing morphological transition due to cluster environmental effects.
\end{abstract}


\section{Introduction}   

Early-type dwarfs are often assumed to be lacking in HI gas and
star formation. Models often picture their star formation history
as a single
burst early in the history of the Universe.  Yet a subset of early-type
dwarfs shows signs of relatively recent star formation, as revealed by
stellar populations, star formation, and gas content (e.g., \citealp{Cons03};
\citealp{Lisk06}, 2007; \citealp{Gross08}).  The detection of recent star
formation in some systems suggests that at least some early-type dwarfs
have formed by transitioning from another galaxy class, e.g., a
later-type spiral or dwarf that loses its gas as it enters the hostile
environment of a galaxy cluster, or perhaps a gas-poor object
accretes gas from another object.  \citet{Lisk07} find that
approximately half of the early-type dwarfs in Virgo belong to an
unrelaxed population that could be associated with recently infalling
galaxies.

Previous studies have attempted to quantify the 
the HI gas content of early-type dwarfs 
(e.g., \citealp{Cons03}), but have been limted by small samples
and heterogeneous observations. 
The Arecibo Legacy Fast ALFA 
(ALFALFA) Survey, a sensitive blind survey of the Arecibo sky
(\citealp{Giov05}), is providing a complete and unbiased view of
HI content and structures in various environments to z=0.06.
Included in the survey is the entire Virgo cluster region, for which
ALFALFA has a limiting HI mass sensitivity of $\sim 2.0 \times 10^7$\msun.

\begin{figure}[!ht]
\begin{center}
\includegraphics[scale=0.3]{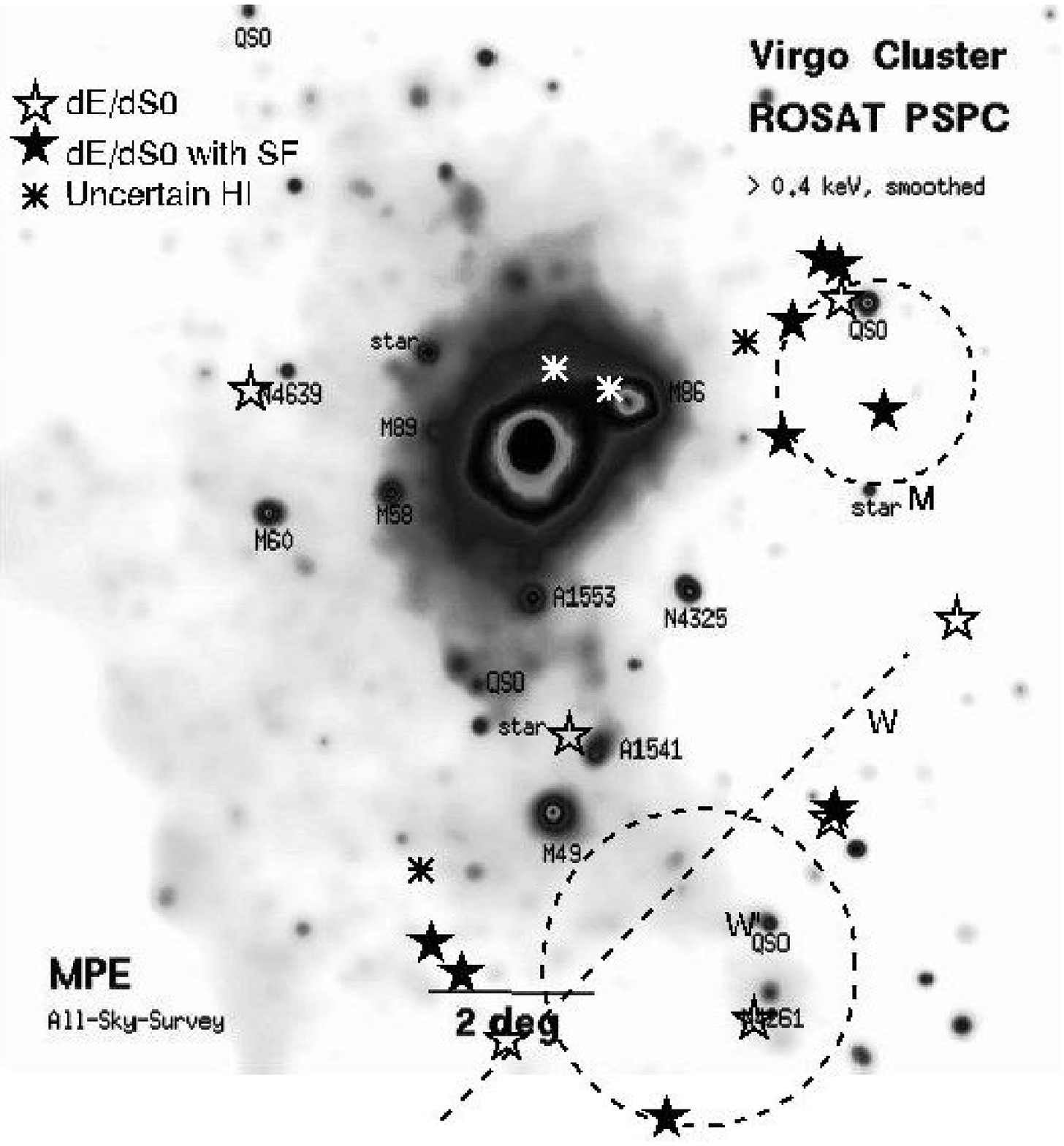}
\includegraphics[scale=0.4]{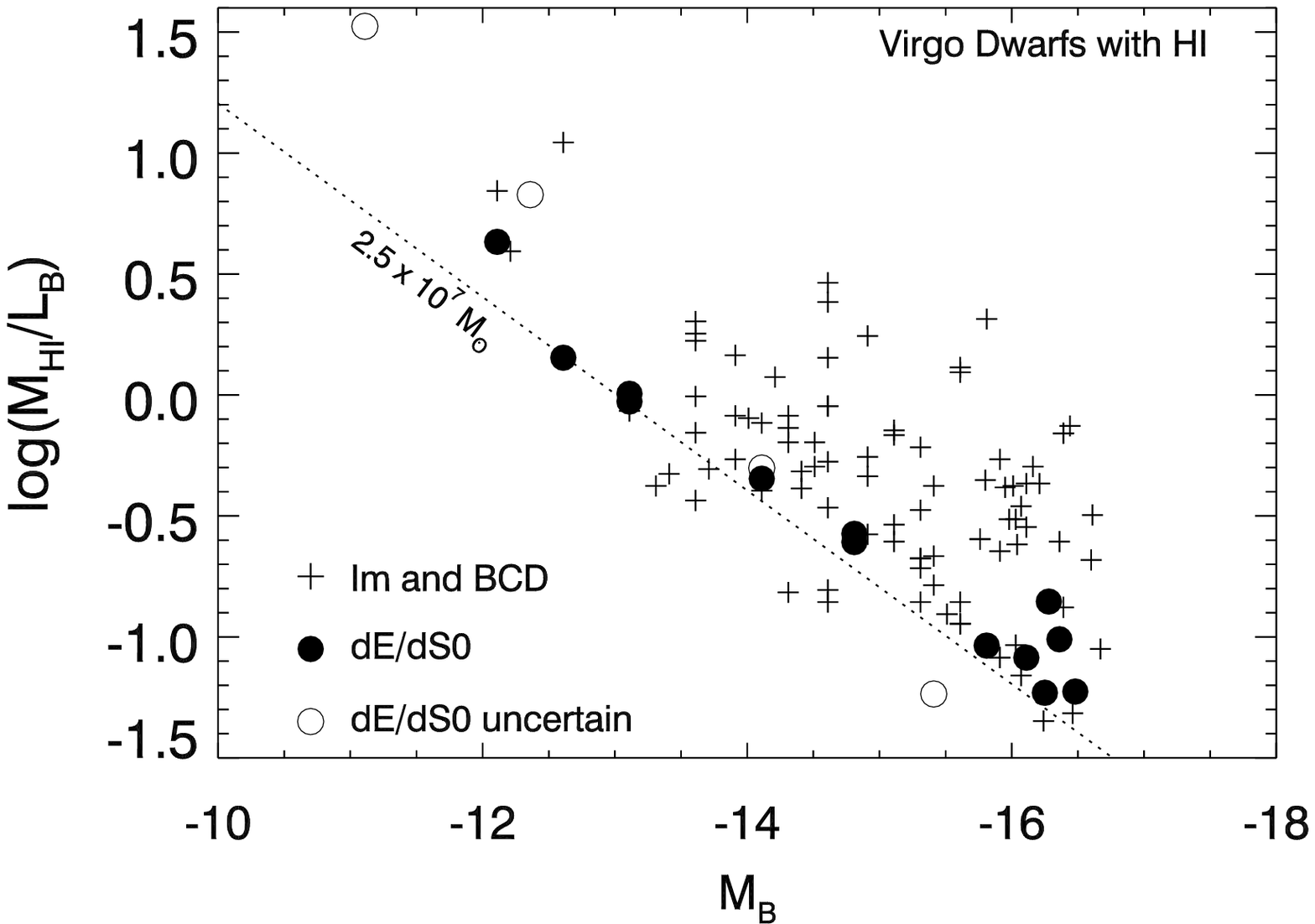}
\end{center}
\caption{
(a) Locations of dE/dS0 galaxies detected by ALFALFA in the declination 
range 4-16 
degrees, superposed on a ROSAT map of the Virgo Cluster 
\citep{Boeh94}.  
The approximate locations of the M, W', and W 
clouds (\citealp{Bing93}) are indicated.
Solid symbols denote dE/dS0 with H$\alpha$ emission.
(b) Log $M_{HI}$/$L_B$ vs $M_B$ for detected dwarfs. 
The dotted line shows the 
completeness limit for a galaxy with an HI mass of $2.5 \times 10^{7}$ \msun~
at the Virgo distance. In both figures, four
dE/dS0 with uncertain HI detections are separately indicated.}
\label{fig1}
\end{figure}

\section{Observations and Results}
\subsection{HI}
ALFALFA HI detections in the Virgo Cluster between the
declinations of 4 - 16 degrees (e.g., \citealp{Giov07}; \citealp{Kent08}) 
were compared with the
Virgo Cluster Catalog (VCC: \citealp{VCC}; \citealp{Bing93})
to identify galaxies classified as dwarfs.  
A total of 17 or 1.6\% of the Virgo dE/dS0 in this
declination range were detected by ALFALFA. 
Four of these are uncertain due to poor 
signal-to-noise and/or interference from M87 or radio noise.
In contrast, Im and BCD galaxies are detected at a rate of 
68\% and 79\%, respectively.

The HI profiles of the detected dE/dS0 are mostly singly-peaked.
Most have velocity widths (uncorrected for inclination) in the range
30-60 \kms.
The median HI mass is 3.5 x 10$^7$\msun. This is 2.2 times smaller 
than the median HI mass of Virgo Im galaxies (and
$\sim$2 times the ALFALFA detection limit at the 
Virgo distance).

Figure~\ref{fig1}a shows the distribution of detected dE/dS0 superposed
on the \citet{Boeh94} ROSAT map of the Virgo Cluster.
Most are located in the cluster outskirts, with several in the 
direction of the M Cloud. 
The median distance from M87 of dE/dS0 with HI is 4.3$^{\circ}$ 
(1.3 Mpc for an assumed Virgo distance of 16.7 Mpc), compared to a 
median distance of 3.2$^{\circ}$ (0.9 Mpc) for the general dE/dS0 population.

The dE/dS0 with HI are 3 times brighter in the median than
the genedominant mechanisms at work will require ral dwarf elliptical population (99.8\% significance). 
This may be partially due to selection effects, since 
fainter dwarfs with a similar fraction of HI per unit luminosity
would fall under
the detection limit for ALFALFA at the distance of Virgo. 
Considering only galaxies with $M_B < -16$, we find that
7\% (5/71) of dE/dS0 are detected.

Figure~\ref{fig1}b shows M$_{HI}$/L$_B$ as a function of M$_B$
for Virgo dwarfs with HI.
The median M$_{HI}$/L$_B$ of dE/dS0 with HI is 0.27, about half
that of Virgo Im or BCD. 
Compared to more isolated dwarfs (not shown), we find that
Virgo dwarfs extend to lower M$_{HI}$/L$_B$. 
%


\subsection{H$\alpha$}
H$\alpha$ images were obtained for 12 early-type dwarfs 
with HI at the CTIO 0.9m (SMARTS),
the Wise Observatory 40-inch, the WIYN 0.9m (courtesy J. Salzer), and from the
Goldmine online database (\citealp{Gav03}). 
75\% (9/12) of early-type dwarfs show H$\alpha$ emission. The median equivalent
width is about 5 times smaller than the value for Im galaxies.
In most cases, the emission is centrally concentrated, extending about
half as far in the median as that of Im galaxies.

\section{Comparison to Previous Results}

\citet{Cons03} reported 7/48 (15\%) Virgo dE/dS0 sample
detected in HI. Six of the detected galaxies are within our 
declination range. Two, the brightest and third-brightest, 
are not detected by ALFALFA and were likely
confused in previous observations. Three detections are reproduced by ALFALFA 
with comparable masses, although one of these, VCC 31, is classified 
\lq ?\rq~in the VCC and is not counted here as dE/dS0. 
A fourth is a marginal detection by ALFALFA requiring 
followup confirmation (not counted as a detection in this
paper).

Lisker \etal~(2006, 2007) present an extensive study of  
413 Virgo dE/dS0 with $M_B < -13.1$, defining sub-types
based on morphological features. We detect 7 of these galaxies
in HI. Four are dE(bc) or dE with blue center, comprising 17\% of the
dE(bc) class. Three of these also show H$\alpha$ emission.
The other four galaxies in the Lisker \etal~sample have four different
sub-type classifications. However, we find no HI detections among 
galaxies classified as dE(di), or dE with disk feature.

\citet{vanZee04}
find that about half of a sample of 16 Virgo dE/dS0 galaxies show
significant rotation, with amplitudes of 20-30 \kms. 
None of the galaxies in their sample is detected in HI by ALFALFA.


\section{Summary}

The ALFALFA survey is providing a homogeneous survey of HI content
in Virgo Cluster dwarf galaxies. Observations to date
reveal that fewer than 2\% of Virgo dE/dS0 dwarfs
have HI reserviors. Those that do typically have
weak H$\alpha$ emission indicating ongoing star formation. 
The H$\alpha$ morphologies  suggest that gas has been
preferentially removed from the outer galaxy and/or funneled into the
central regions. Virgo dwarfs with HI classified as dE/dS0 could be 
transitional objects, evolving from gas-rich progenitors 
subjected to cluster environmental effects. 
Aperture synthesis observations will be helpful in 
distinguishing dominant mechanisms.

\acknowledgements 
The conference organizers and
the support of NSF AST-0724918 (RAK) are gratefully acknowledged.


\end{document}